\documentclass[12pt,preprint]{aastex}
\begin{document}

\title{The evolution of X-ray emission in young stars}

\author{Thomas Preibisch\altaffilmark{1} \and
Eric D. Feigelson\altaffilmark{2}}

\altaffiltext{1}{Max-Planck-Institut f\"ur Radioastronomie,
 Auf dem H\"ugel 69, D-53121 Bonn, Germany}
\altaffiltext{2}{Department of Astronomy \& Astrophysics,
Pennsylvania State University, University Park PA 16802}

\slugcomment{Version: \today}

\begin{abstract}

The evolution of magnetic activity in late-type stars is part of the 
intertwined rotation-age-activity relation which provides an
empirical foundation to the theory of magnetic dynamos.  We study 
the age-activity relation in the pre-main sequence (PMS) regime, for 
the first time using mass-stratified subsamples.  The effort is 
based on the $Chandra$ Orion Ultradeep Project (COUP) which provides 
very sensitive and homogenous X-ray data on a uniquely large sample 
of 481 optically well-characterized low-extinction low-mass members 
of the Orion Nebula Cluster, for which individual stellar masses and 
ages could be determined. More than 98 percent of the stars in this 
sample are detected as X-ray sources.

Within the PMS phase for stellar ages in the range $\sim 0.1-10$ 
Myr, we establish a mild decay in activity with stellar age 
$\tau$ roughly as $L_{\rm X} \propto \tau^{-1/3}$.  On longer 
timescales, when the Orion stars are compared to main sequence 
stars, the X-ray luminosity decay law for stars in the $0.5 < M < 
1.2$~M$_\odot$ mass range is more rapid with $L_{\rm X} \propto 
\tau^{-0.75}$ over the wide range of ages $5 < \log \tau < 9.5$ yr.  
When the fractional X-ray luminosity $L_{\rm X}/L_{\rm bol}$ and the 
X-ray surface flux are considered as activity indicators, the decay 
law index is similarly slow for the first $1-100$ Myr but 
accelerates for older stars.  The magnetic activity history for M 
stars with masses $0.1 < M < 0.4~M_\odot$ is distinctly 
different. Only a mild decrease in X-ray luminosity, and even a mild 
increase in $L_{\rm X}/L_{\rm bol}$ and $F_{\rm X}$, is seen over 
the $1-100$ Myr range, though the X-ray emission does decay over 
long timescales on the main sequence.

Together with COUP results on the absence of a
rotation-activity relation in Orion stars, we find that the
activity-age decay is strong across the entire history of
solar-type stars but is not attributable to rotational
deceleration during the early epochs.  A combination of
tachocline and distributed convective dynamos may be
operative in young solar-type stars.  The results for the
lowest mass stars
are most easily understood by the dominance of convective
dynamos during both the PMS and main sequence phases.

\end{abstract}


\section{Introduction \label{intro.sec}}

The magnetic activity of late-type stars during their main
sequence (MS) phase has long been known to be correlated
with rotation and anti-correlated with age
\citep{Skumanich72}.  This age-rotation-activity relation
has been well-documented in the X-ray band which traces
coronal and flare processes associated with the eruption of
magnetic fields from the stellar interior onto the surface.
Studies of young open clusters have documented a decay of
average X-ray luminosities by two orders of magnitude from
their arrival onto the Zero Age Main Sequence (ZAMS) to
ages of several tenths of a Gyr \citep[see reviews
by][]{Randich97, Jeffries99, Micela01, FavataMicela03}. The
decay continues through the later $1-10$ Gyr epochs
\citep{Guedel97, Feigelson04}.

The commonly accepted explanation for this phenomenon is
that late-type stellar magnetic activity is regulated
principally by rotation \citep{Pallavicini81, Noyes84,
Baliunas95}, and its decay with stellar age is attributed
to rotational breaking due to mass loss.  Ionized wind
particles gain high specific angular momentum as they
travel outward along spiral-shaped magnetic field lines
that corotate with the star \citep{Schatzman62, Kawaler88}.
Slower rotation presumably leads to reduced velocity shear
at the tachocline  between the radiative and convective
zones, resulting in reduced magnetic field generation by
the $\alpha-\Omega$ dynamo \citep{Schrijver00}.

X-ray studies of low-mass pre-main sequence (PMS)
stars\footnote{X-ray emission is the most widely available
tracer of surface magnetic activity for pre-main sequence and
young main sequence stars
\citep{Feigelson99, FavataMicela03}. Chromospheric emission
from optical and ultraviolet spectroscopic line indicators,
such as H$\alpha$ or the Mg II lines, are often overwhelmed
by emission attributed to accretion onto the surface.
Nonthermal radio emission from flares can only occasionally
be detected. In comparison, X-rays produced mainly by
magnetic reconnection flares are seen in virtually all PMS
stars across the Initial Mass Function and at all early
phases of evolution from Class I protostars through the
ZAMS \citep{Getman05b, Preibisch05}.} have revealed very
strong activity, exceeding the solar levels by several
orders of magnitude both in time-averaged luminosity and
power of individual flares \citep{Feigelson99,
FavataMicela03, Guedel04}. However, PMS activity does not
exhibit the relationships to rotation and age seen along
the MS. First, the rotational evolution is complicated by
competition between an acceleration due to stellar
contraction, a deceleration probably due to magnetic
coupling between the outer layers and the circumstellar
disk, and an uncertain coupling between inner and outer
layers within the star \citep[e.g.][]{Bouvier97,
Krishnamurthi97, Barnes03a, Wolff04}. Second, the
correlation between magnetic activity measured by X-ray
luminosity and surface rotation is absent in PMS stars
\citep{Feigelson03, Flaccomio03a, Preibisch05}. This may
either arise because the $\alpha-\Omega$ dynamo is
saturated or, as these stars are often fully convective
without a tachoclinal layer, because magnetic field
generation is dominated by a turbulent convective dynamo
that does not depend on rotation. Recent research thus
establishes that the age-rotation and rotation-activity
relations can not be readily extrapolated from the MS into
the PMS regime.

We investigate here the third bivariate connection seen in MS stars: 
the age-activity relation.  Past evidence for a decay in magnetic 
activity between $10^5-10^7$~yr before the star has begun to burn 
hydrogen has been unclear and ambiguous.  Consider, for example, the 
stellar population of the Chamaeleon I cloud.  From an {\it Einstein 
Observatory} study, \citet{Feigelson89} reported an order of 
magnitude enhancement in the X-ray luminosity function (XLF) over 
that seen in the ZAMS Pleiades cluster. However, this was found 
to be wrong by the {\it ROSAT} study of the Chamaeleon I cloud which 
showed that {\it Einstein} sources were often poorly resolved blends 
of several closely spaced PMS stars and that the sample of 
weak-lined T Tauri stars was badly incomplete \citep{Feigelson93, 
Lawson96}. The resulting XLF from this study showed little or no 
enhancement over ZAMS levels. Studies of portions of the extended 
Taurus-Auriga clouds indicated little decline in the X-ray 
luminosity function between the ages of PMS and Pleiades stars 
\citep{Walter91, Briceno97} although an enhancement in flare 
luminosities among PMS stars was noted \citep{Stelzer00}. 
\citet{Kastner97} compared average X-ray luminosities for a variety 
of young stellar clusters studied with {\it ROSAT} and found an 
increase of the median fractional X-ray luminosities by about one 
order of magnitude during the first 100~Myr, followed by a steep 
decay for older ages.

The contradictions among these early studies were likely
due to combinations of several problems: different levels
of incompleteness due to sensitivity limitations;
incomplete cluster samples, particularly among X-ray faint
weak-lined T Tauri stars; inadequate spatial coverage of
nearby star formation regions subtending large regions of
the sky; inadequate spatial resolution to resolve crowded
regions; and inadequate hard energy sensitivity to
penetrate into embedded regions.  The confounding influence
of a strong statistical association between X-ray
luminosity and mass \citep{Preibisch05} is particularly
important: clusters studied with lower sensitivity tend to
derive higher mean X-ray luminosities due to the failure to
detect lower mass stars, leading to spuriously high levels
of magnetic activity \citep{Feigelson99, PZ02,
Feigelson03}.

These difficulties are largely overcome with the 
\dataset[ADS/Sa.CXO#obs/COUP]{the {\it Chandra\/} Orion 
Ultradeep Project (COUP)}.
It achieves high sensitivity $\log 
L_{\rm X} \geq 27.0$ erg s$^{-1}$ with complete spatial coverage of 
the lightly obscured Orion Nebula Cluster (ONC) with spatial 
resolution $\geq 500$ AU \citep{Getman05a}. This gives a large 
unbiased sample of young stars down to the stellar limit;  the 
$L_{\rm X}-$Mass relation indicates it is $>$95\%-complete X-ray 
detections above $\simeq 0.1$ M$_\odot$ \citep{Preibisch05}.
Furthermore, the sample is so large and well-characterized from 
optical study \citep{Hillenbrand97} that the age-activity relation 
can be examined in mass-stratified subsamples.  The Orion Nebula 
field also exhibits a range of ages within the PMS phase 
\citep{Palla99}, allowing the study of X-ray evolution between $\sim 
0.1-10$ Myr. COUP thus provides a unique opportunity to reexamine 
the PMS age-activity relationship in a reliable and quantitative 
fashion.

\section{The COUP data}

The COUP observation is the deepest and longest X-ray
observation ever made of a young stellar cluster, providing
a rich and unique dataset for a wide range of science
studies.  All observational details and a complete
description of the data analysis can be found in
\citet{Getman05a}. The total exposure time of the COUP image
is 838\,100 sec (232.8 hours or 9.7 days), and an source
detection procedure located 1616 individual X-ray sources
in the COUP image. The superb point spread function and the
high accuracy of the aspect solution allowed a clear and
unambiguous identification of most X-ray sources with
optical or near-infrared counterparts with median offsets
of 0.2\arcsec.  The remaining sources are a mixture of new
PMS stars and unrelated extragalactic background sources
\citep{Getman05b}.

Spectral analysis was performed to produce an acceptable 
spectral model and to give reliable time-averaged broadband 
luminosities for as many as possible sources. A complete and 
detailed description of the spectral analysis can be found in 
\citet{Getman05a}, here we only summarize the main points. The XSPEC 
spectral fitting programme was used to fit the extracted spectra 
with one- or two-temperature optically thin thermal plasma MEKAL 
models \citep{Mewe91}, assuming 0.3 times solar abundances  
\citep{Imanishi01,Feigelson02a} and X-ray absorption according to 
the atomic cross sections of \citet{Morrison83} with traditional 
solar abundances to infer a total interstellar column density.

We note that, in general, the resulting spectral parameters and 
their uncertainties are often not reliably determined, and 
alternative models may be similarly successful. In these 
cases, preference was given to the solution that avoids the 
inference of a very luminous, heavily absorbed, ultra-soft ($kT_1 < 
0.5$ keV) component. The study presented in this paper, however, is 
not strongly affected by these uncertainties. The sample used for 
our analysis is restricted to optically visible T~Tauri~stars with 
modest extinction ($A_V \leq 5$~mag, corresponding to $\log N_H 
\lesssim 22$~cm$^{-2}$), which are not affected by the greatest 
uncertainties in plasma temperatures and emission measures that 
occur for sources with high absorption, $\log N_H \simeq 22-23$ 
cm$^{-2}$.

The intrinsic, extinction-corrected, X-ray luminosities of the 
sources were computed by integrating the best-fit model source flux 
over the $[0.5\!-\!8]$ keV band. The typical uncertainties in the 
derived X-ray luminosities of the stars in our sample are $\lesssim 
30\%$. Note that these X-ray luminosities represent the average over 
the 10 days exposure time of the COUP dataset. This implies that the 
effect of short excursions in the X-ray lightcurves, like flares 
with typical timescales of a few hours, are ``smoothed out''. 
The detection limit of the COUP data is $L_{\rm X,min} \sim 
10^{27.0}$~erg/sec for lightly absorbed stars, and few sources 
appear near that limit. 

We use the tabulated X-ray properties (and upper limits for the 
undetected ONC members) and identifications of the COUP sources 
listed in Tables 8, 9, and 11 in \citet{Getman05a}. Throughout this 
paper, we use the absorption-corrected X-ray luminosity in the 
$0.5-8$ keV band, denoted $\log L_{t,c}$ by \citet{Getman05a}, for 
which  we use here the simpler appellation, $\log L_{\rm X}$. 


\section{Definition of the ONC low-mass star sample}

The basis for the construction of the sample of well-characterized 
ONC stars used here is the study by \citet{Hillenbrand97} [H97 
hereafter] of $1576$ optically visible ($I < 17.5$) stars within 
$\sim 2.5$~pc ($\sim 20'$) of the Trapezium, for $934$ of which 
optical spectral types are known. We used an updated version of the 
H97 tables in which spectral types and other stellar parameters for 
many objects have been revised \citep[see][]{Getman05a}. H97 discuss 
that, while their optical database is missing very low mass and 
heavily obscured objects, it is representative of all stars in the 
ONC region. 

In this study, we use a homogenous, optically selected and 
extinction limited sample of low-mass ONC members. Our sample 
consists of those ONC stars from H97 which are: (i) located within 
the field-of-view of the COUP observation; (ii) not classified as 
unrelated fore- or background stars on the basis of their proper 
motions; (iii) for which the optical extinction is known and is $A_V 
\le 5$~mag; and (iv) for which masses and ages could be 
determined by comparison with the PMS models of \citet{Siess00}. A 
detailed discussion of these selection criteria is given in 
\citet{Preibisch05}. We further restrict our analysis to 
low-mass stars with $M \leq 2\,M_\odot$. While the vast majority of 
these stars are clearly detected as X-ray sources in the COUP data, 
a few stars remained undetected. Most of the non-detections are due 
to X-ray source confusion in the COUP data; the typical case are 
close ($\sim 1''-2''$ separation) binary systems, in which only one 
of the components is clearly detected as an X-ray source 
\citep{Getman05b}. In these cases, the object would perhaps have 
been detected if located at a different position. Since the 
occurrence of source confusion should not depend on stellar 
parameters, we consider these objects to be ``unobserved'' and 
ignore them in our analysis. 

These criteria provide a sample of 481 ONC stars, 474 of which 
are detected as X-ray sources. With an X-ray detection fraction of 
98.5\%, we can be very confident that our conclusions will not 
be affected by non-detections. While our optical sample is not 
complete because spectral types are not available for all ONC 
stars, it should be a statistically representative sample of the 
ONC young stellar population with low extinction. The only potential 
systematic selection effect might be that some of the older 
($\gtrsim 10$~Myr) very-low mass ($M \lesssim 0.2\,M_\odot$) stars, 
such as reported by \citet{Slesnick04}, may be missing.


\section{Evolution of X-ray activity within the age range of the ONC}

\subsection{ONC stellar age estimates}

The large number of PMS stars with individual age estimates in our 
COUP sample provides an unique opportunity to look for possible 
relations between the X-ray activity and stellar age. The HR diagram 
for the ONC shows a wide spread of $L_{\rm bol}$ values at any given 
$T_{\rm eff}$ value, suggesting a considerable spread of stellar 
ages. The ages of the individual stars in our COUP optical sample 
derived from comparison of the HR diagram to the PMS models of 
\citet{Siess00} range from $\log\left(\tau\,[{\rm yr}]\right) = 2.7$ 
up to $\log\left(\tau\,[{\rm yr}]\right) = 8.4$. 

These values, however, have to be treated with caution. Inferred 
ages less that $\sim 0.1$ Myr are unreliable due to inadequate 
theoretical treatment of the stellar birth process.  The extent of 
age spread between $\sim 0.1$ to $\sim 10$ Myr for the ONC stars has 
been debated \citep{Hillenbrand97, Palla99, Slesnick04}. 
Difficulties include choice of sample stars, choice of evolutionary 
tracks, and the empirical determination of the bolometric 
luminosities and effective temperatures. Bolometric luminosities 
have several potential sources of errors including: intrinsic 
photometric variability, which can be very large for accreting 
systems; error in the estimation of the extinction, which is can be 
affected by circumstellar material; and the possible presence of 
unresolved binary companions, which causes systematic overestimation 
of the luminosity both because the two components light is 
attributed to the primary component and because the reddening is 
overestimated by a redder secondary. Unresolved binary companions 
can lead to overestimates of the luminosity by factors of two, what 
can lead to systematic underestimations of the true age up to factor 
of four.  Errors in the determination of stellar effective 
temperatures affect the age estimates directly for the more massive 
($M > 1\,M_\sun$) stars, and indirectly (via errors in the amount of 
de-redenning and bolometric corrections) for the less massive $M < 
1\,M_\sun$ stars. We will therefore refer to the individual stellar 
age values derived from the observed HR-diagram as ``isochronal 
ages" to make clear that they are not necessarily identical to the 
true stellar ages.

For the $M < 1\,M_\sun$ stars we estimate that variability and 
extinction errors will typically cause an artificial spread in 
stellar luminosities of $\Delta \log L \sim \pm 0.1$, while the 
unresolved binary companions can cause a spread of $\Delta \log L 
\sim + 0.3$. One might thus expect a perfectly coeval sample of 
stars to exhibit a scatter of $\Delta \log L \sim {+0.4 \choose 
-0.1}$ around the isochrone in the observed HR diagram. The observed 
scatter in the HR diagram of the ONC is considerably larger than 
this with $\Delta \log L \sim 1.5$.  We therefore concur with 
\citet{Palla99} and \citet{Slesnick04} that a true age spread of up 
to about a factor of 100 (from $\sim 0.1$ to $\sim 10$~Myr) is 
present in the ONC.

\subsection{The dependence of X-ray emission on
age within the ONC \label{Lx_age_ONC.sec}}

With these caveats in mind, we can now look for the relations 
between X-ray activity and stellar (isochronal) age.  As the 
evolutionary time scales for PMS stars are a strong function of the 
stellar mass \citep{Preibisch05}, which itself has dependencies on 
bolometric luminosity, we believe it is necessary to consider mass 
stratified subsamples to avoid confusing interrelationships. We 
therefore consider four separate mass ranges: $0.1-0.2\,M_\odot$, 
$0.2-0.4\,M_\odot$, $0.4-1\,M_\odot$, and $1-2\,M_\odot$.

We also consider three measures of X-ray emission: the X-ray 
luminosity of the star, $\log L_{\rm X}$ (in units of erg s$^{-1}$); 
the fraction of the bolometric luminosity emitted in the X-ray band, 
$L_{\rm X}/L_{\rm bol}$ (in dimensionless units); and the X-ray 
surface flux, $F_{\rm X}$ (in units of erg $^{-1}$ cm$^{-2}$). The 
fractional X-ray luminosity and X-ray surface flux are very similar 
to each other within a mass stratum, as they differ by $T_{\rm 
eff}^4$ and PMS stars on the Hayashi convective tracks with 
similar masses also have similar surface temperatures. However, 
conceptually these are different indicators:  the fractional 
luminosity tells the overall efficiency of the star in converting 
luminous energy into magnetic activity, while the surface flux gives 
the average magnetic activity emissivity at the stellar surface.

The distribution function of these X-ray emission measures are not 
completely defined by the data because of the 7 undetected stars 
(all of which are in the two lower mass strata).  To account for the 
non-detections, we construct the Kaplan-Meier maximum-likelihood 
estimator of the X-ray distribution functions.  Similarly, for 
comparing two samples, seeking bivariate correlations and 
calculating linear regressions, we use statistical methods developed 
for `survival analysis' which treat the non-detections in 
mathematically correct fashions \citep{Feigelson85}.  These 
calculations were performed with the {\it ASURV}\footnote{Astronomy 
SURVival analysis \citep{LaValley90} available from StatCodes 
\url{http://www.astro.psu.edu/statcodes}} package. The probabilities 
that a correlation is present are calculated using a generalization 
of the nonparametric Kendall's tau statistic. The linear regressions 
are calculated using the EM (Expectation-Maximization) Algorithm 
under the assumption of Gaussian residuals in the ordinate around 
the fitted line. The results of the statistical analysis are 
summarized in Table~\ref{fit.tab}.

\begin{figure}
\plotone{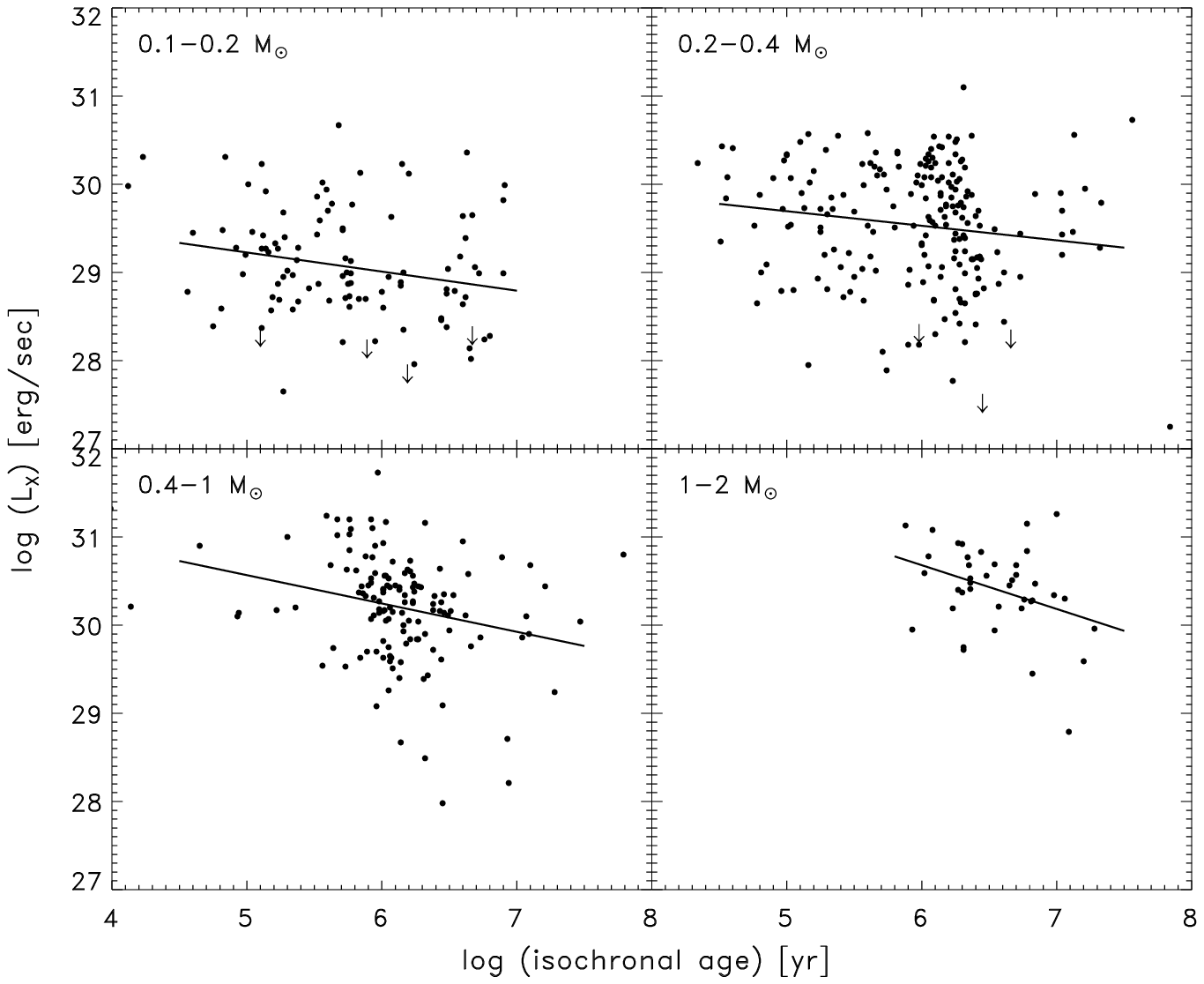}
\caption{X-ray luminosity versus stellar age for the stars
in the COUP ONC sample separated in
four mass ranges. The lines shows the linear regression
fits computed with the EM Algorithm. \label{lx_age.fig}}
\end{figure}

\begin{figure}
\plotone{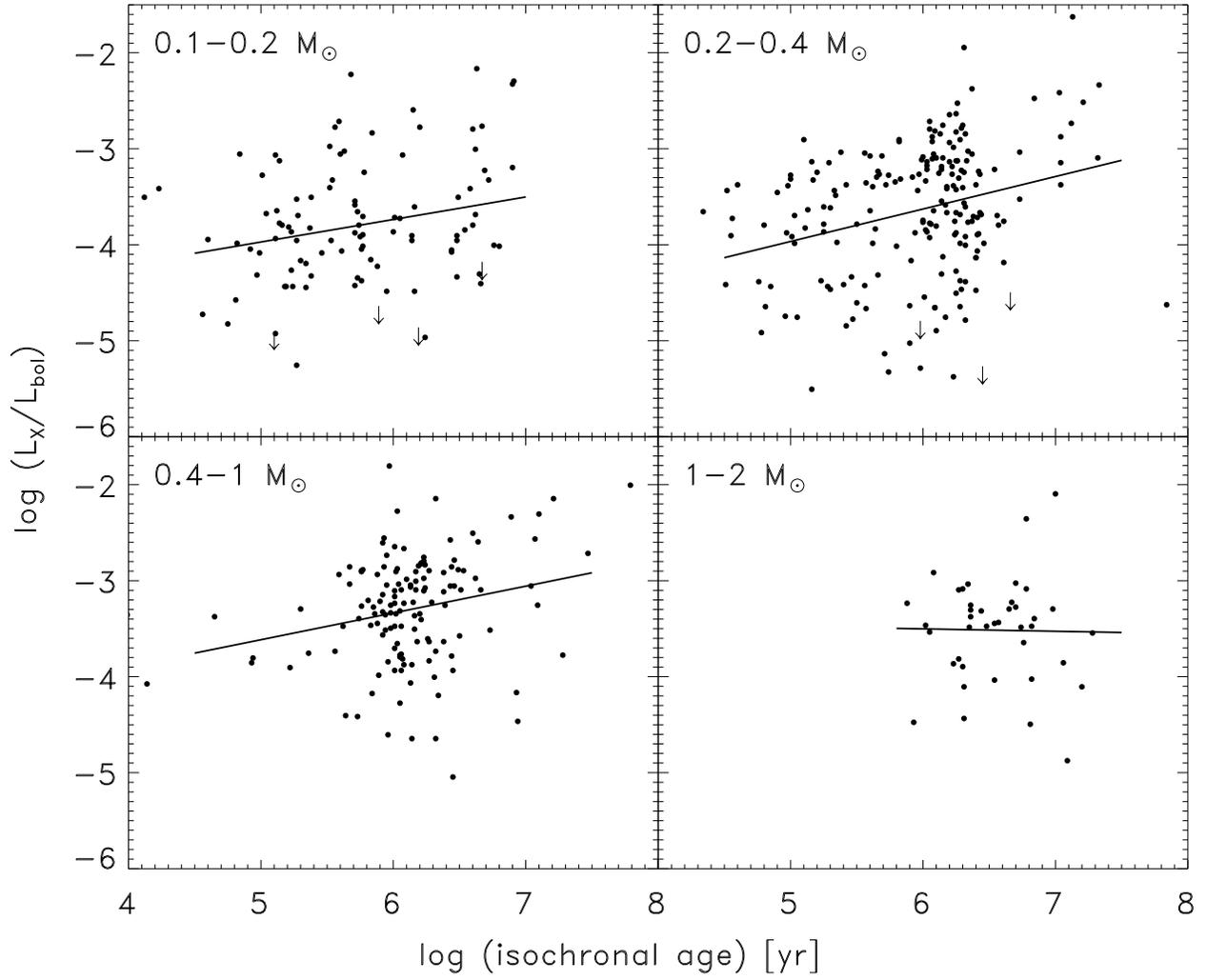}
\caption{Fractional X-ray luminosity versus stellar age.
Stellar sample and fits are as described in Figure
\ref{lx_age.fig}. \label{lxlb_age.fig}}
\end{figure}

\begin{figure}
\plotone{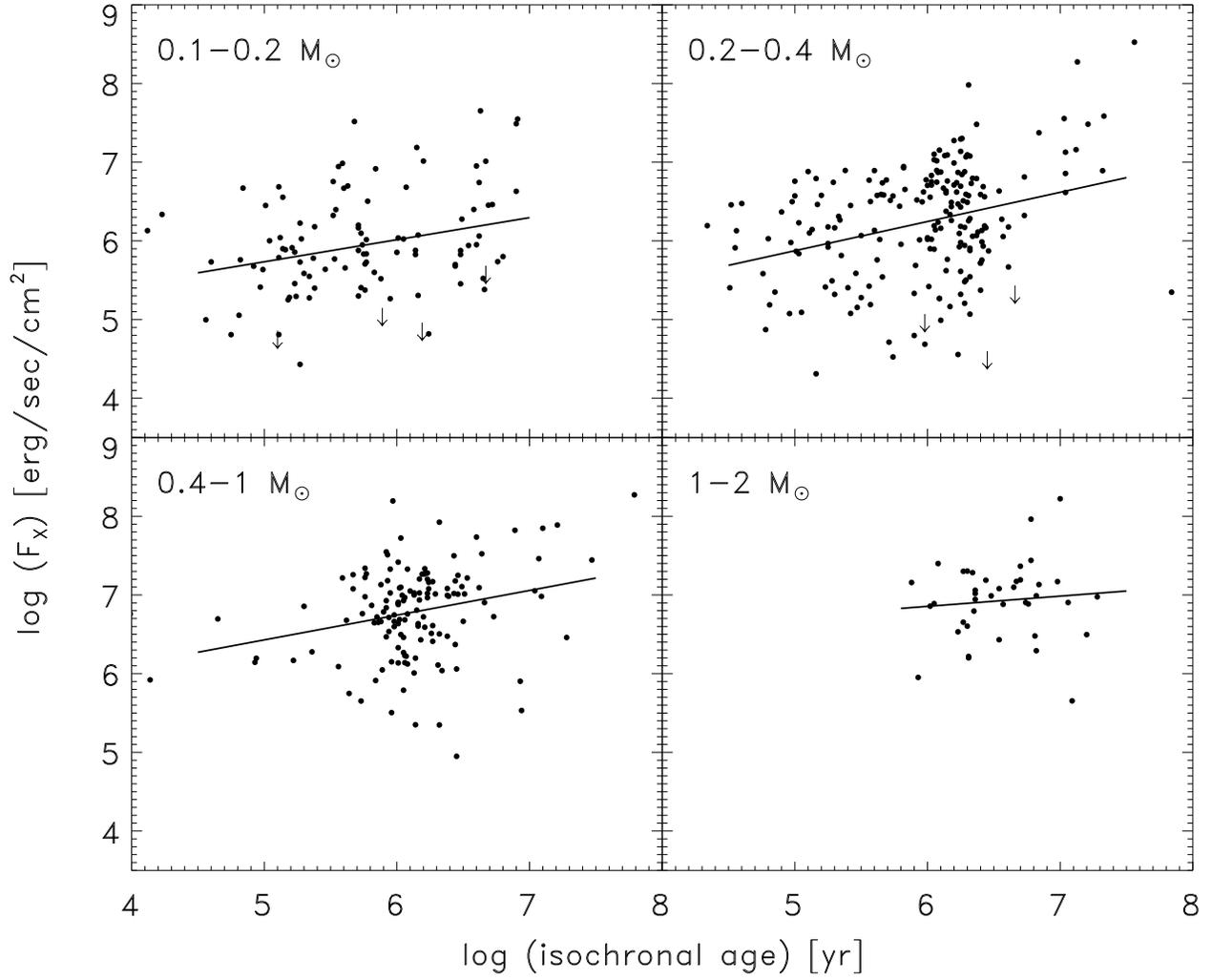}
\caption{X-ray surface flux versus stellar age.  Stellar
sample and fits are as described in Figure
\ref{lx_age.fig}. \label{fx_age.fig}}
\end{figure}

Figures~\ref{lx_age.fig}--\ref{fx_age.fig} plot the X-ray indicators 
against isochronal stellar ages $\tau$ for our lightly absorbed 
optical sample. The X-ray luminosity (Figure~\ref{lx_age.fig}) is 
seen to decrease with isochronal age from $\sim 0.1$ to $\sim 10$ 
Myr in all four mass ranges, though the effect is not strong. For 
the $L_{\rm X}$ $vs.$ isochronal age diagram, the correlation tests 
in ASURV give probabilities ranging from 95\% to 99.92\% in the four 
mass strata that an anticorrelation is present. The slopes of the 
linear regression fits to the $\log\left(L_{\rm X} \right)  = a + 
b \times  \log\left(\tau\right)$ relations with the EM algorithm 
range between $b \sim -0.2$ and $b \sim -0.5$\footnote{Since 
it is now well-established that X-ray luminosities of accreting PMS 
stars are systematically lower than non-accretion PMS stars 
\citep{Flaccomio03c, Preibisch05}, one might expect that $L_{\rm 
X}$ might increase with age rather than decrease as more accreting 
classical T Tauri stars evolve into non-accreting weak-lined T Tauri 
stars.  We find that this effect does not dominate the stronger 
X-ray decay we find here.  There are two reasons why this effect is 
minor: the population of accreting stars is relatively small 
compared to non-accreting stars in the ONC sample; and the 
difference in X-ray luminosities of the two classes is only $\Delta 
\log L_{\rm X} \simeq 0.4$ which is relatively small compared to the 
full range of the XLFs.}.

The fractional X-ray luminosity (Figure~\ref{lxlb_age.fig}) shows an 
increase with isochronal age for the low-mass stars, but with lower 
statistical significance. The slopes of the linear regression fits 
to the $\log\left(L_{\rm X}/L_{\rm bol}\right)  = a + b  \times  
\log\left(\tau\right)$ relations with the EM algorithm are about 
$b\sim 0.3$ for the lower mass strata. The apparent decrease of 
$L_{\rm X}/L_{\rm bol}$ with isochronal age in the $1-2\,M_\odot$ 
mass range is not statistically significant.

The X-ray surface flux (Figure~\ref{fx_age.fig}) also increases with 
isochronal age for the low-mass stars with correlation probabilities 
above 99\% confidence for the low-mass ($< 1\,M_\odot$) stars.  The 
slopes of the linear regression fits to the $\log\left(F_{\rm 
X}\right)  = a + b \times  \log\left(\tau\right)$ relations with 
the EM algorithm are $b \sim 0.3$. Again, no significant effect 
is found for the $1-2\,M_\odot$ stars.


\section{Long term evolution of X-ray activity}

We now consider the temporal evolution of X-ray properties over a 
longer timescale by comparing the ONC PMS stars to samples of 
late-type stars in older star formation regions, the Pleiades, the 
Hyades and the solar neighborhood.  We wish to trace the evolution 
of stars with fixed mass which, when comparing PMS and ZAMS stars, 
implies different spectral types.  For example, according to the 
models of \citet{Siess00}, a $1\,M_\odot$ star at the age of 1~Myr 
will have a spectral type of K5 rather than G2 on the main sequence. 
To match past treatments of this question, we consider here mass 
strata corresponding to the ZAMS spectral types: mass range 
$0.1-0.5$~M$_\odot$ corresponding to spectral types M0--M5; 
$0.5-0.9$~M$_\odot$ corresponding to K0--K9; and $0.9-1.2$~M$_\odot$ 
corresponding to G0--G9.

\subsection{The comparison samples of stars}

From Figures~\ref{lx_age.fig}-\ref{fx_age.fig}, we adopt an average 
age of 1 Myr for the ONC sample.  We have selected for comparison 
two, somewhat older, PMS stellar clusters which have been 
well-studied with the $ROSAT$ satellite: NGC~2264, with an adopted 
age of 1.7~Myr; and the Chamaeleon I star forming region with an 
adopted age 5.5~Myr \citep{Flaccomio03b}.  Due to sensitivity 
limitations, these samples are not available for the lowest mass 
stars.

To characterize young main sequence stars, we adopt the X-ray and 
optical data for the Pleiades and Hyades samples obtained from 
\citet{Stelzer01}. These samples are based on $ROSAT$ observations 
of optically selected members and therefore include upper limits for 
X-ray undetected stars. The Pleiades sample consists of 41 G-type 
stars (including 18 upper limits), 112 K-type stars (including 41 
upper limits), and 62 M-type stars (including 27 upper limits). The 
Hyades sample consists of 22 G-type stars (including 2 upper 
limits), 51 K-type stars (including 6 upper limits), and 90 M-type 
stars (including 32 upper limits).  We assume ages of 80~Myr for the 
Pleiades\footnote{We note that recent age estimates for the Pleiades 
give a considerably higher age, perhaps 130 Myr. The uncertainty 
does not affect the conclusions drawn in this paper.}
 and 650~Myr for the Hyades.

For older Galactic disk stars, we use the NEXXUS database
\citep{Schmitt04} provides updated $ROSAT$ X-ray and optical data 
(including accurate $HIPPARCOS$ parallaxes) for a well-defined 
volume-limited samples of nearby field stars. The samples have 43 
G-type stars (including 5 non-detections) within a limiting distance 
$d_{\rm lim} = 14$~pc, 54 K-type stars (including 2 non-detections) 
within $d_{\rm lim} = 12$~pc, and 79 M-type stars (including 5 
non-detections) within $d_{\rm lim} = 6$~pc.  The NEXXUS tables were 
kindly provided to us by the authors; they list $M_V$, $B-V$, the 
X-ray luminosity $L_{\rm X}$  and the X-ray surface flux $F_{\rm 
X}$.  We calculate bolometric luminosities by interpolation from the 
main sequence relationship between $M_V$ and $L_{\rm bol}$. Although 
these stars possess a wide range of ill-determined ages, we adopt an 
average age of 3 Gyr for this sample.

\subsection{Construction of the X-ray distribution functions}

When comparing X-ray luminosities determined with different X-ray 
observatories, one has to take into account the different energy 
bands for which X-ray luminosities were computed. For Pleiades, 
Hyades, and nearby field stars considered here, the X-ray 
luminosities in the literature are given for the $0.1-2.4$~keV 
$ROSAT$ band. Conversion into the $Chandra$ $0.5-8$~keV band for 
comparison with our COUP results depends on the X-ray spectrum. We 
calculate the conversion factor using the PIMMS tool. For the 
moderately active ZAMS Hyads and Pleiads, we scale the $ROSAT$ 
luminosities downward by $0.14$~dex corresponding to a plasma 
temperatures $T = 10$~MK and consistent with the count-rate to 
luminosity transformation factor used by \citet{Stelzer01}. For the 
NEXXUS stars, the count-rate to luminosity transformation factor 
used by \citet{Schmitt04} is valid for a plasma temperature of $T 
\sim 2.5$~MK, and the corresponding $ROSAT-Chandra$ band correction 
factor is $-0.33$~dex. The X-ray distribution functions, including 
the X-ray luminosity function (XLF), are the Kaplan-Meier estimators 
introduced in \S~\ref{Lx_age_ONC.sec}.

\subsection{Results}

\begin{figure}
\plottwo{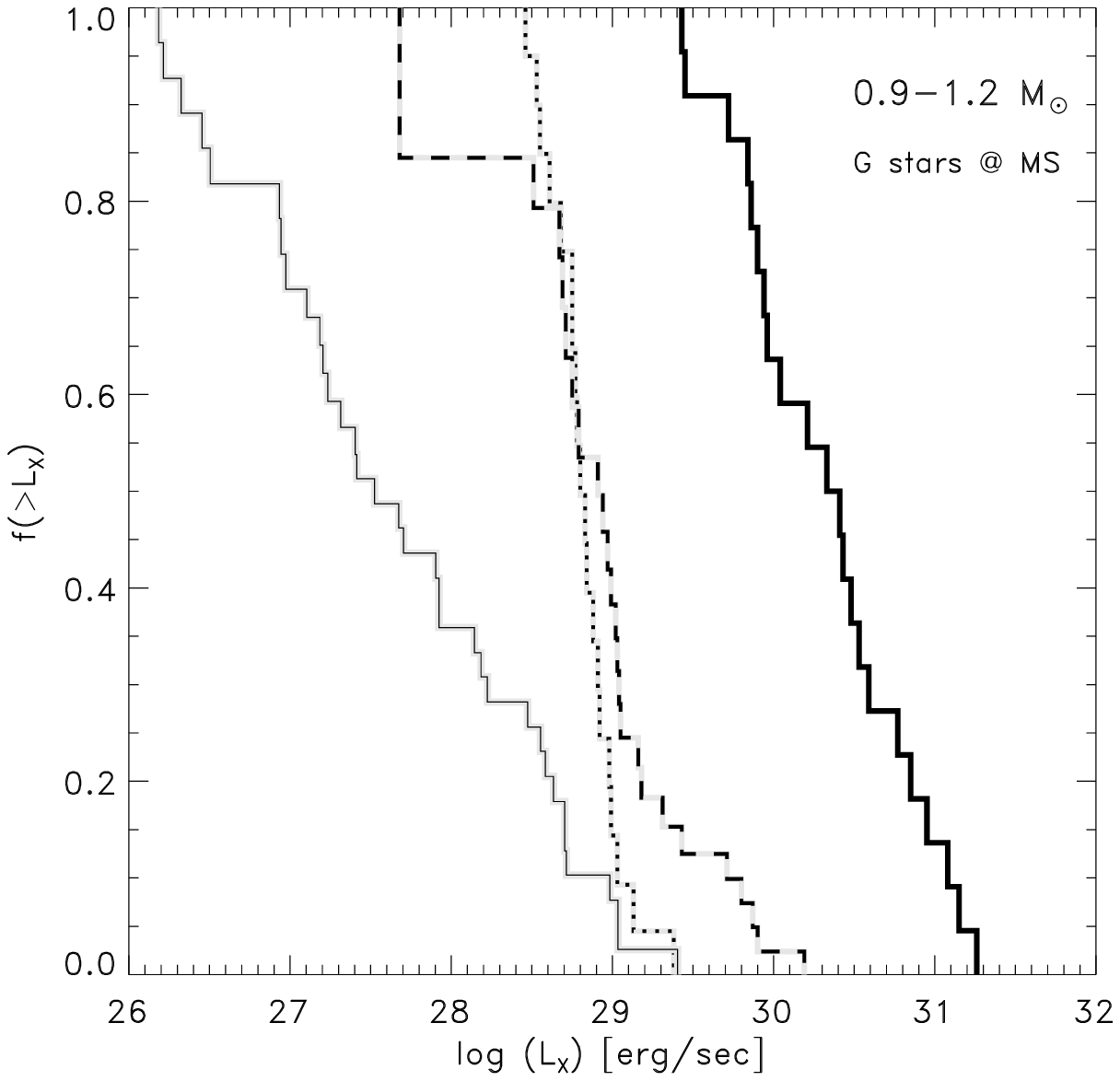}{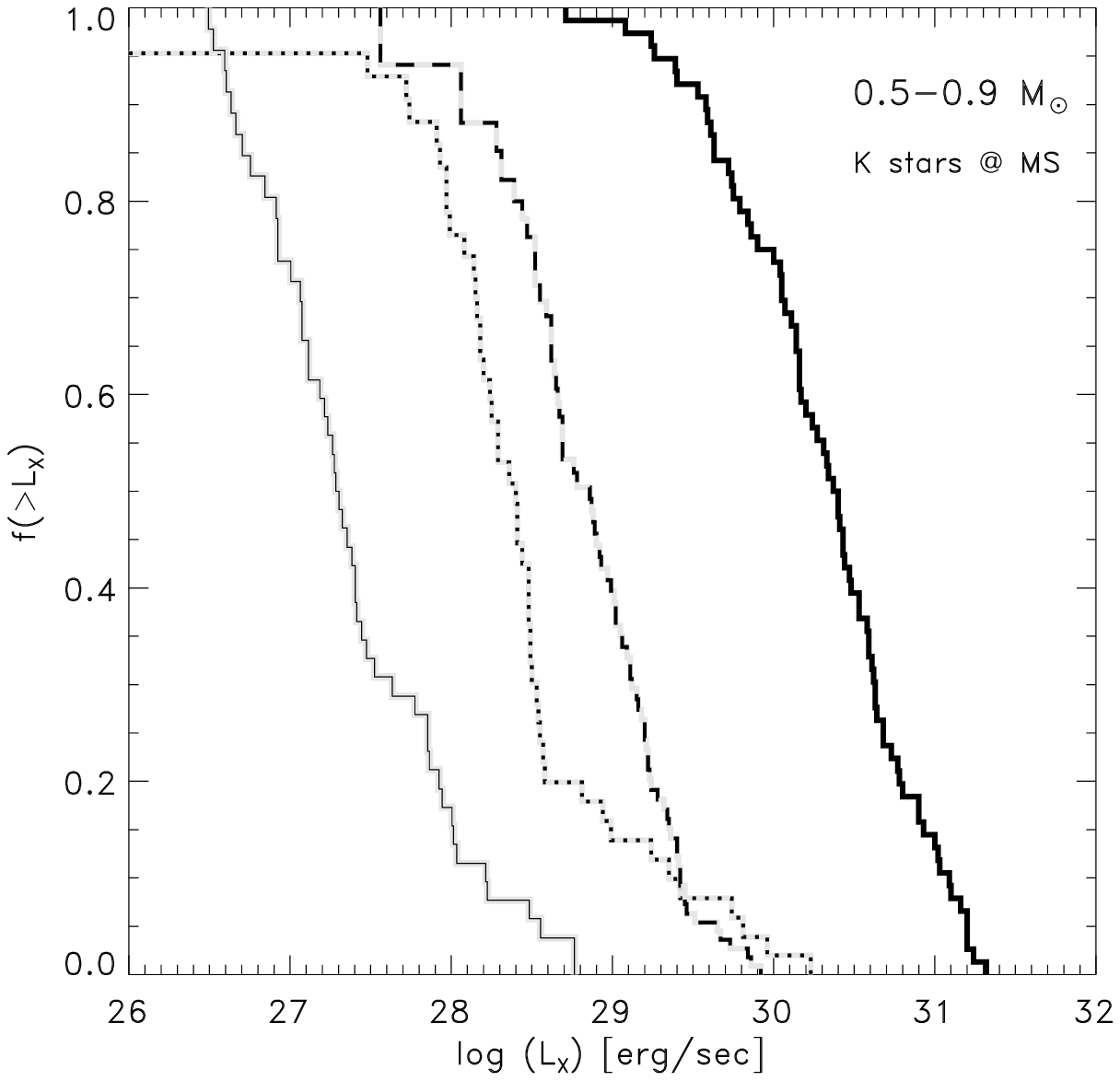} \plottwo{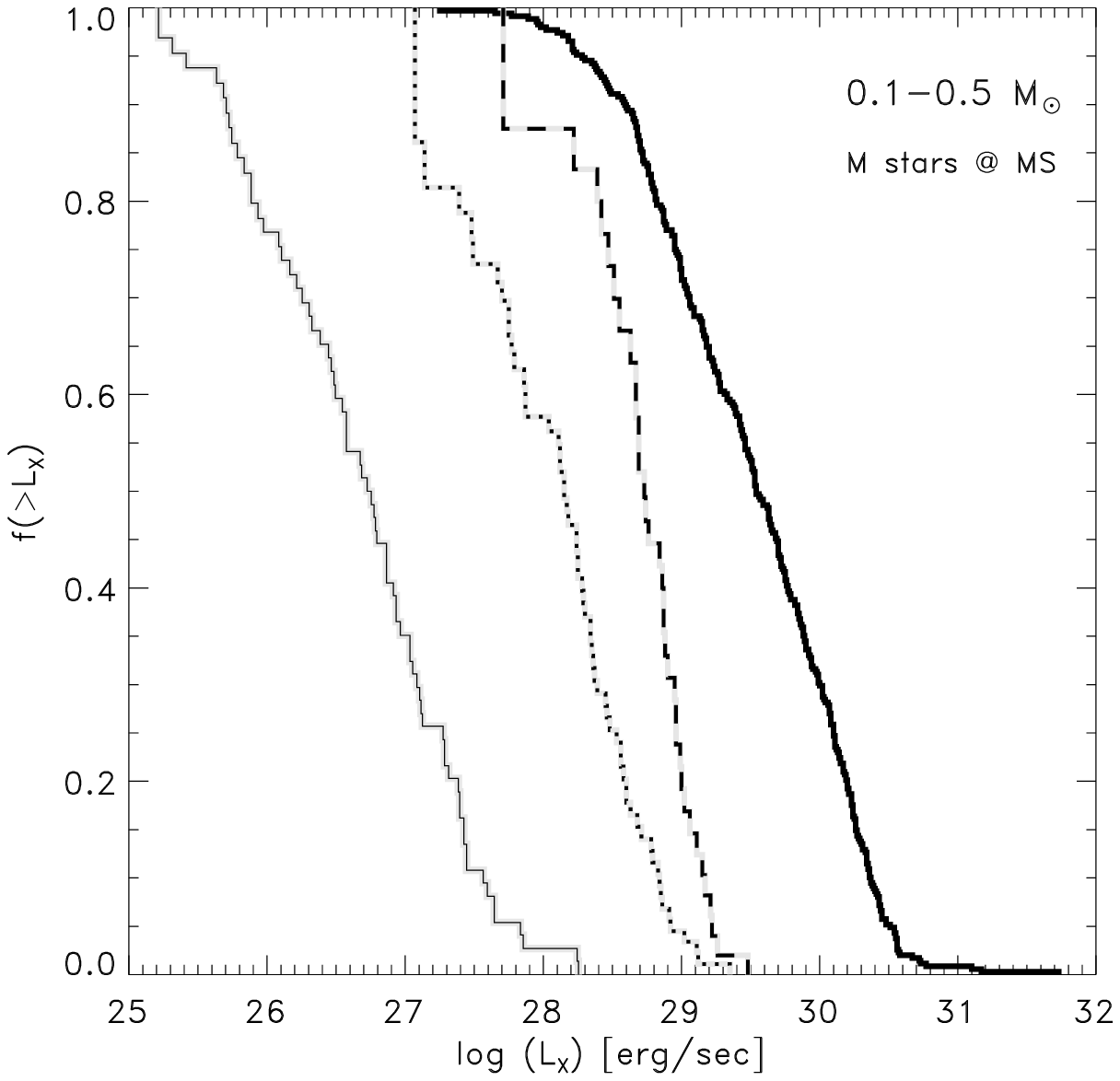}{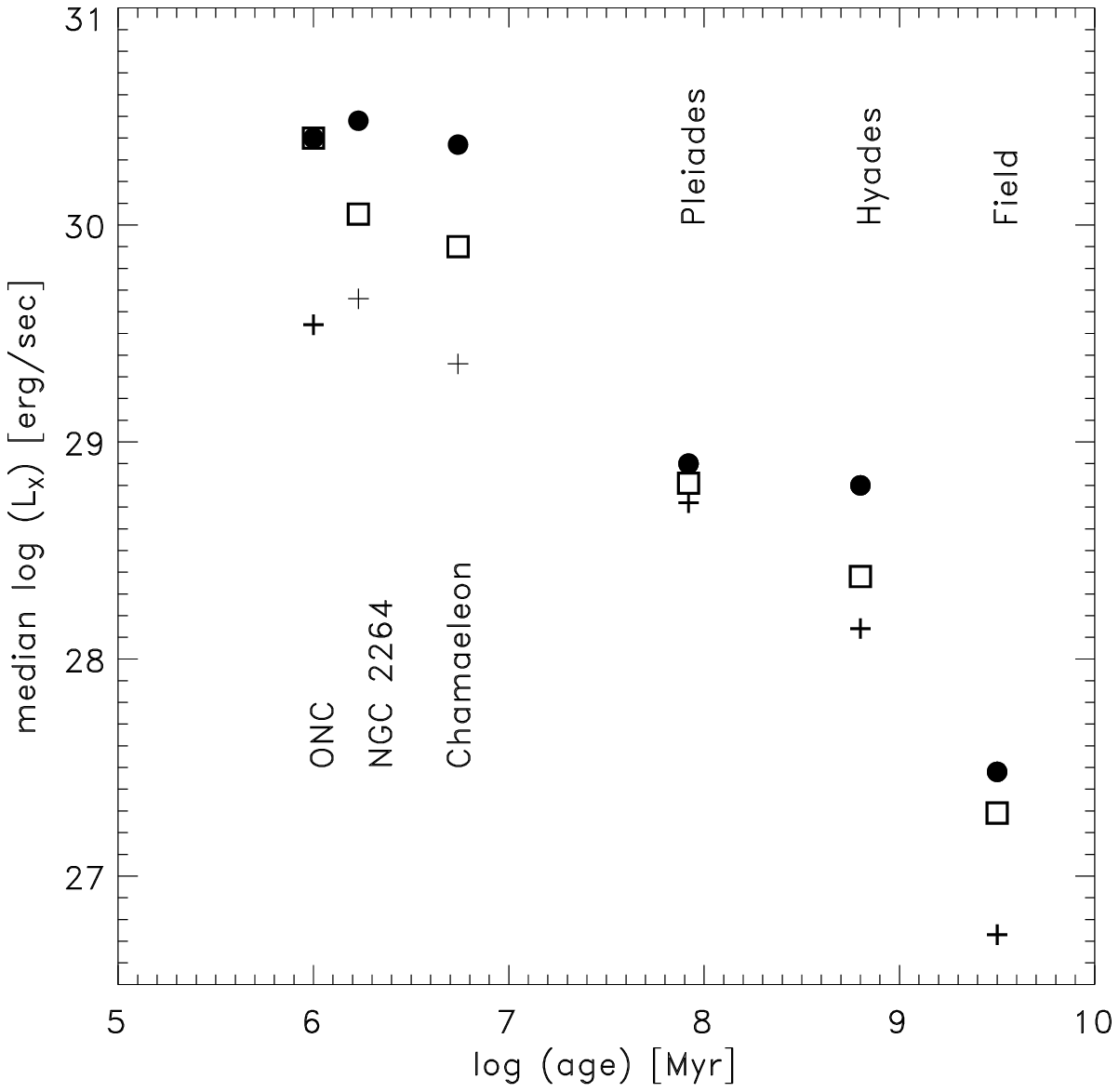} 
\caption{Evolution of the X-ray luminosity from the pre-main 
sequence through the main sequence. The first three plots show the 
cumulative X-ray luminosity functions (XLFs) for three mass ranges.  
In each panel, the thick solid line shows the COUP X-ray 
luminosities for our ONC sample, the dashed line shows the Pleiades, 
the dotted line the Hyades, and the thin solid line the field stars 
from NEXXUS. The lower right panel shows the median X-ray luminosity 
in each sample and mass range as a function of the age.  The solid 
dots show the $0.9\!-\!1.2\,M_\odot$ (G-type on the main sequence) 
stars, squares show the $0.5\!-\!0.9\,M_\odot$ (K-type on the main 
sequence) stars, and crosses the $0.1\!-\!0.5\,M_\odot$ (M-type on 
the main sequence) stars. \label{cdf_lx.fig}}
\end{figure}

\begin{figure}
\plottwo{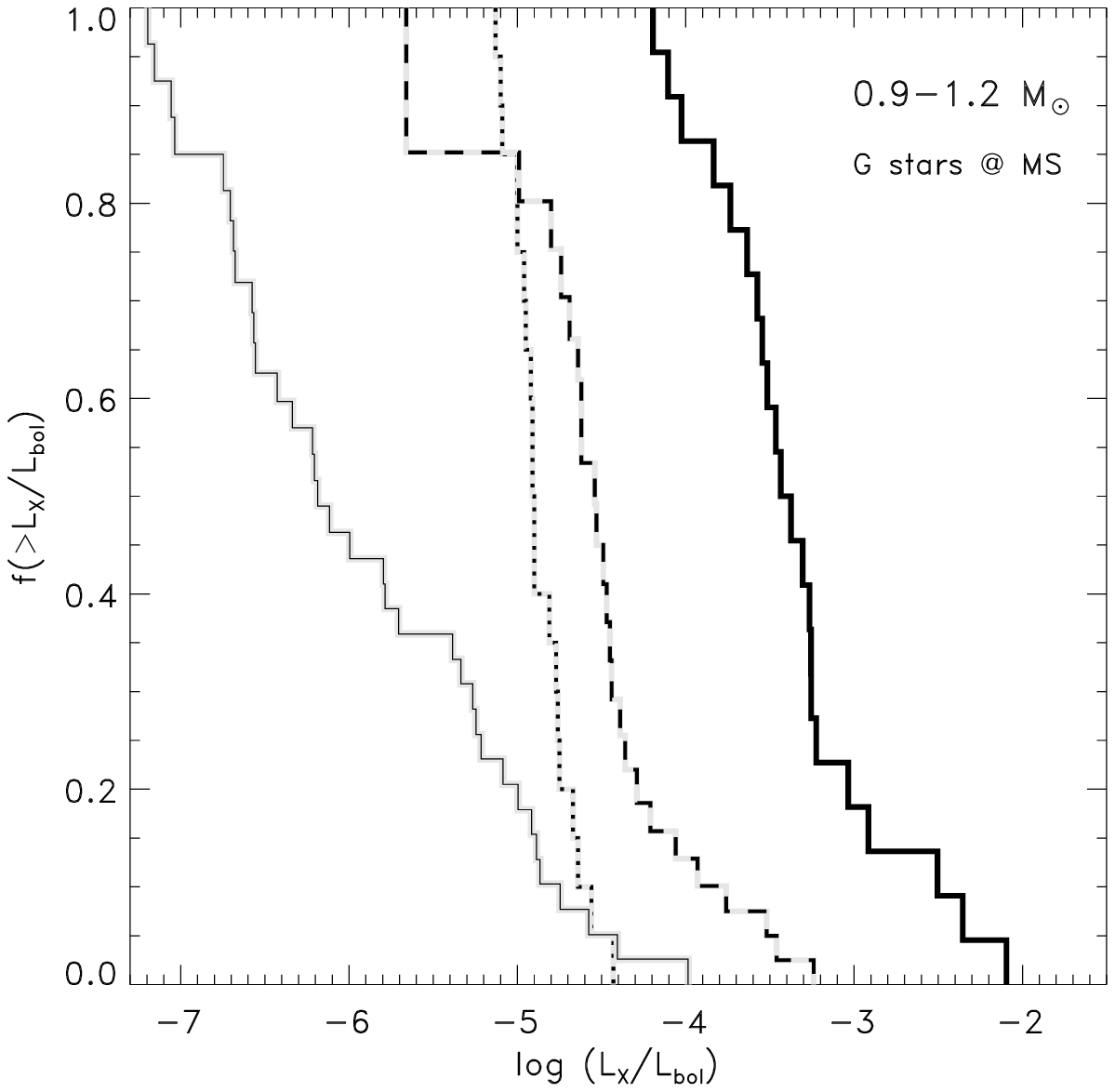}{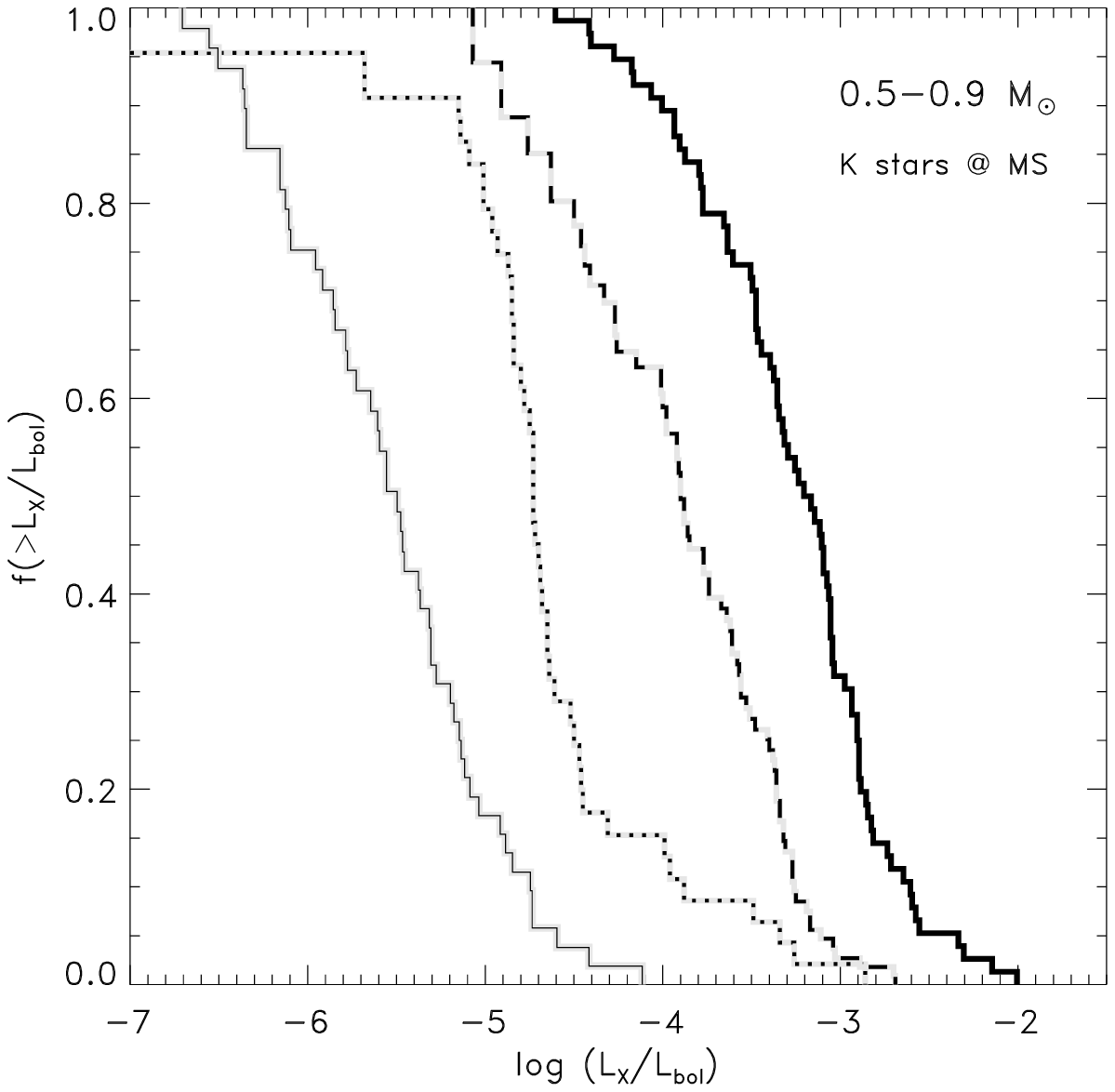} \plottwo{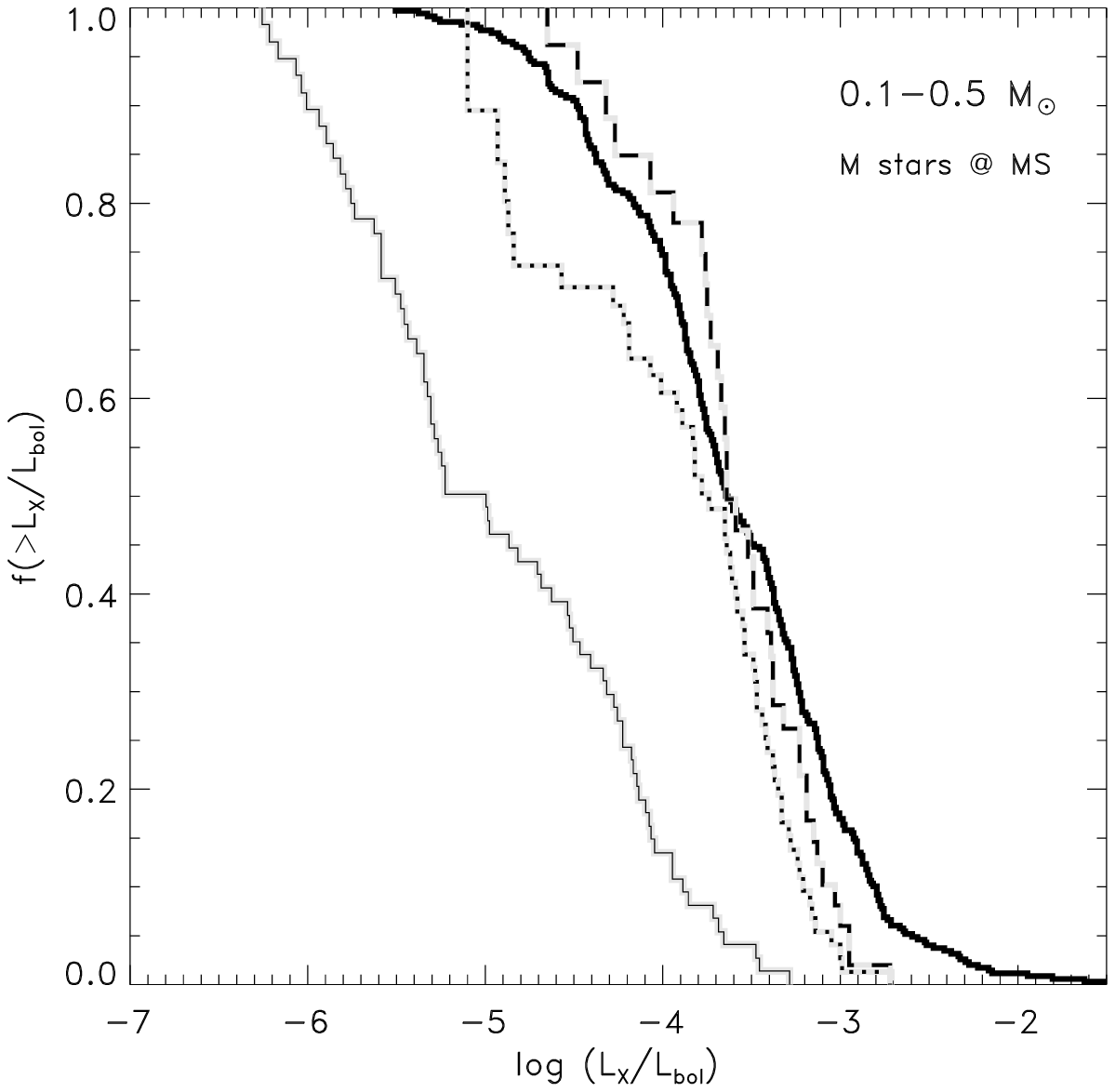}{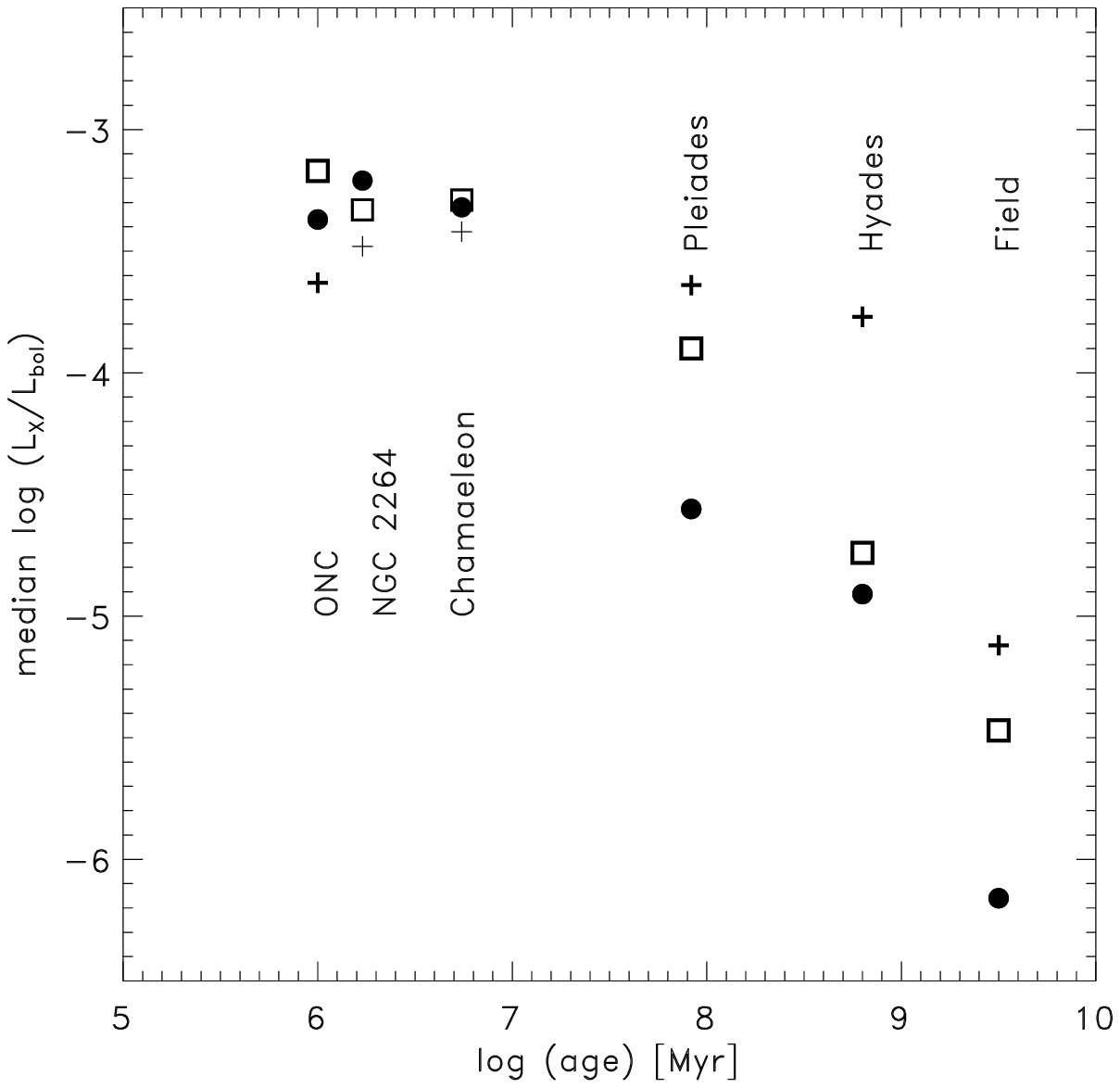} 
\caption{Evolution of the fractional X-ray luminosity. The lines 
and symbols are defined in the Figure~\ref{cdf_lx.fig} caption. 
\label{cdf_lxlb.fig}}
\end{figure}

\begin{figure}
\plottwo{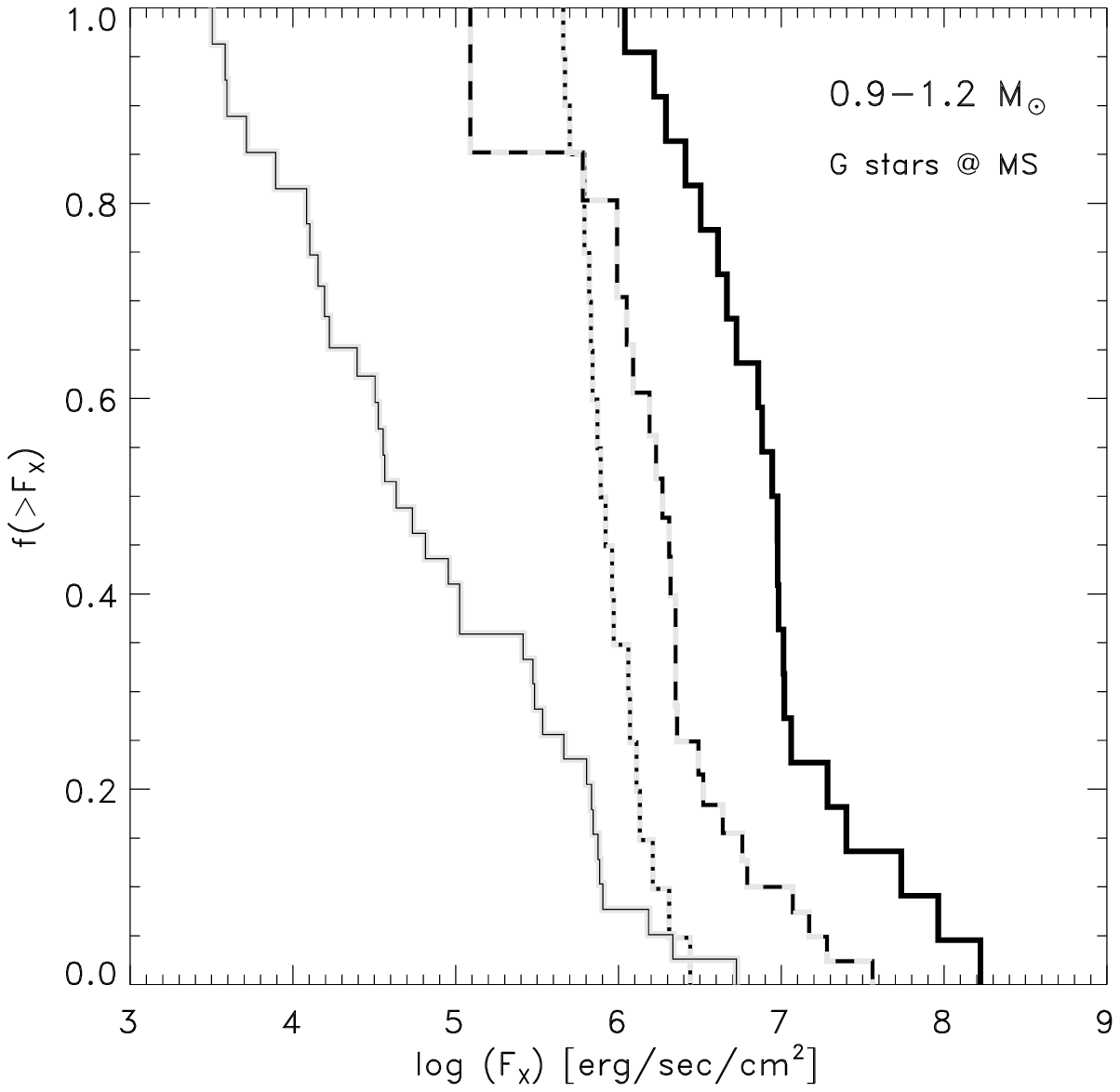}{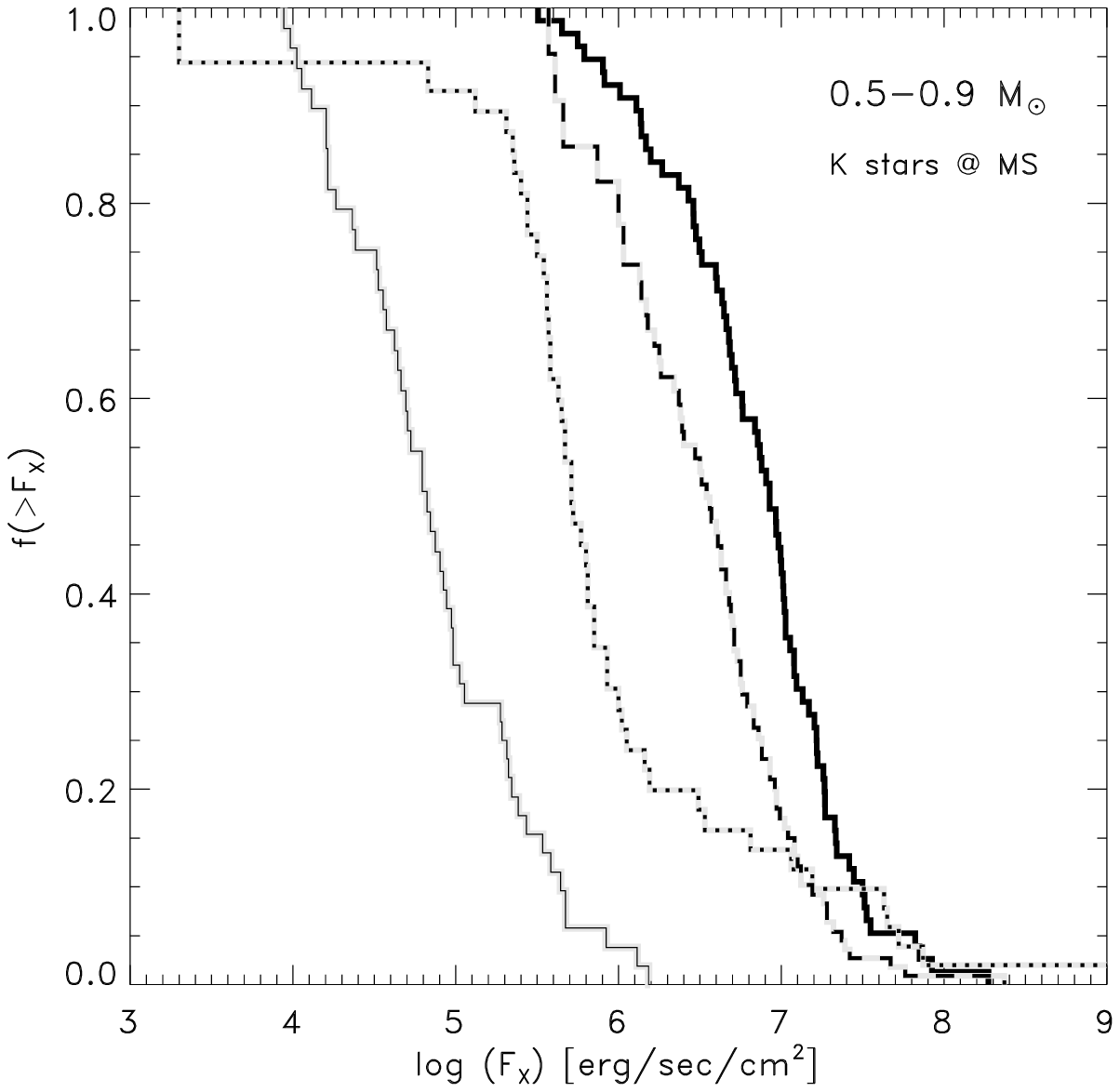} \plottwo{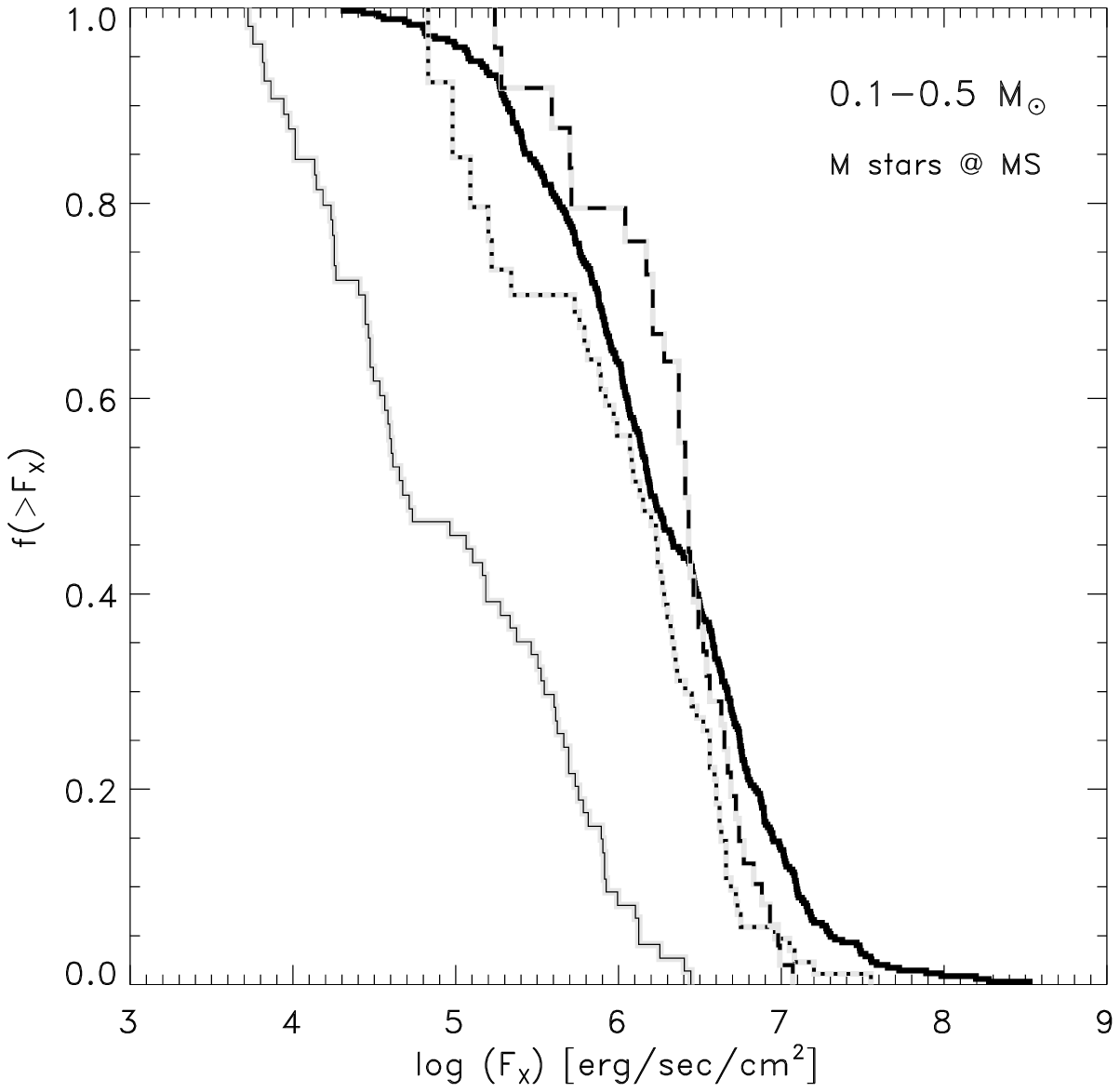}{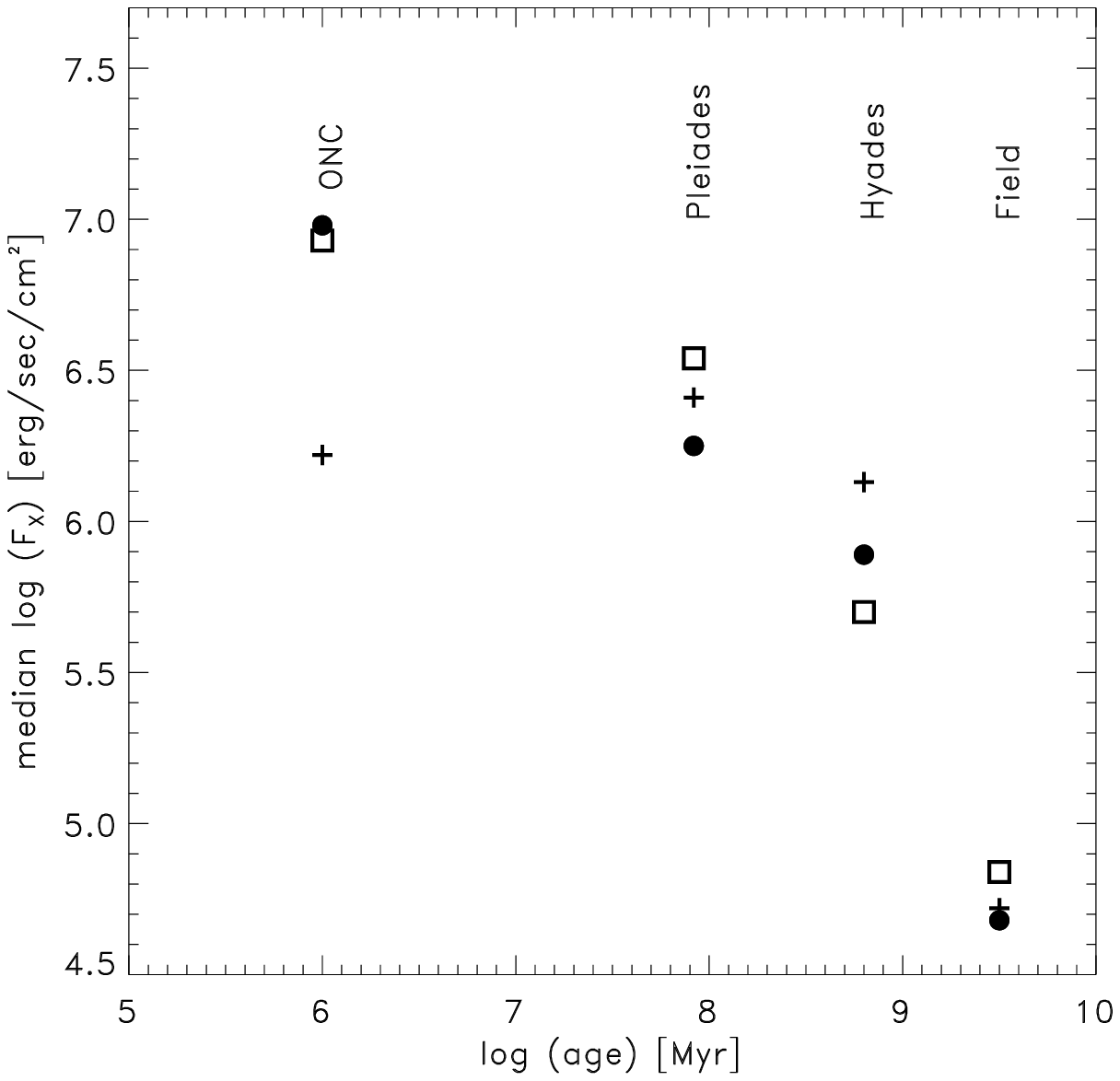} 
\caption{Evolution of the X-ray surface flux. The lines and 
symbols are defined in the Figure~\ref{cdf_lx.fig} caption. 
\label{cdf_fx.fig}}
\end{figure}

Figures \ref{cdf_lx.fig}--\ref{cdf_fx.fig} show the cumulative 
distribution functions of $L_{\rm X}$, $L_{\rm X}/L_{\rm bol}$, and 
$F_{\rm X}$ for the ONC stars based on the COUP results in  
comparison with the older PMS samples, the ZAMS Pleiades and Hyades, 
and the solar neighborhood NEXXUS stars.  The bottom right panels 
compare the median values of the X-ray activity in these samples as 
a function of stellar age.

The X-ray luminosities of Orion stars, in all of the mass strata, 
are highly elevated with respect to all main sequence stars.  For 
$\simeq 1$~M$_\odot$ stars, for example, the ONC sample is $\sim 30$ 
times stronger in X-ray luminosity than Pleiads or Hyads 
(Figure~\ref{cdf_lx.fig}, top left).  The fractional X-ray 
luminosities are elevated for stars $M>0.5$~M$_\odot$ but not for 
lower mass (main sequence M-type) stars (Figure~\ref{cdf_lxlb.fig}).

If we ignore the behavior within the PMS phase and consider only the 
longer-term decline from the ONC to the Pleiades, Hyades and nearby 
field star samples, a rough powerlaw decay in magnetic activity is 
seen for X-ray luminosities. For a relationship of the form 
$\log\left(L_{\rm X}\,[{\rm erg/sec}]\right) = a + b  \times  
\log\left(\tau\,[{\rm Myr}]\right)$, we find slopes of  $b=-0.76$ 
for the mass range $0.9-1.2$~M$_\odot$, $b=-0.78$ for 
$0.5-0.9$~M$_\odot$, and  $b=-0.69$ for $0.1-0.5$~M$_\odot$. The 
decay in the other activity indicators, $L_{\rm X}/L_{\rm bol}$ and 
$F_{\rm X}$, is less rapid with $b \simeq -0.5$ for the higher-mass 
stars and $b \simeq -0.3$ for the $0.1-0.4$~M$_\odot$ stars.  But 
the relationships here do not not appear linear in the log-log 
diagrams of Figures \ref{cdf_lxlb.fig}-\ref{cdf_fx.fig}.  Rather the 
decay starts slowly and accelerates, though not as steeply as an 
exponential decay law as suggested by \citet{Walter91}.

We also note that a global decay law for the whole age range 
($10^6 - 10^{10}$~yr) may not imply a single physical mechanism 
here. Up to ages of a few 10 Myr, X-ray activity does {\em not} 
depend on the stellar rotation \citep{Preibisch05}, the 
stellar interiors for most stars are fully convective, and the stars 
contract. On their approach to the main-sequence, the stars develop 
radiative cores,  which probably enable  $\alpha-\Omega$ type 
dynamo action. Finally, for the young main sequence stars a 
rotation-activity relation is well established 
\citep[e.g.][]{Pizzolato03} and the subsequent decay in X-ray 
activity can be understood as the result  of the wind-driven 
spin-down process. Therefore, it appears rather likely that the 
processes that determine the decline in X-ray activity at 
very young ages are different from those at ZAMS and later stages. 

While the X-ray activity is clearly decreasing on long timescales, 
the evolution within the PMS epoch is less rapid.  Our intra-ONC 
analysis (\S~\ref{Lx_age_ONC.sec}) shows a decay with powerlaw index 
$b \simeq -0.3$ \footnote{Comparison with the NGC~2264 and 
Chamaeleon I clusters suggests there may be no decay at all during 
the PMS phase (Figure~\ref{lx_age.fig}), but there is some question 
whether these $ROSAT$ surveys have sufficient resolution and 
sensitivity for reliable comparison with COUP.  The intra-ONC 
analysis with uniformly excellent $Chandra$ data suggest that the 
decay results in \S~\ref{Lx_age_ONC.sec} should be reliable.}. This 
decay is sufficiently slow that fractional X-ray luminosity and the 
X-ray surface flux is roughly constant or even {\em increases} over 
the age range $0.1-10$~Myr as the stars contract. Such a 
temporary increase in $L_{\rm X}/L_{\rm bol}$ has been previously 
reported from an inter-cluster $ROSAT$ study by \citet{Kastner97}.


\section{Discussion}

The COUP observation provides the most sensitive, uniform
and complete study of X-ray properties for a PMS stellar
population available to date.  The extensive optical
spectroscopy by \citet{Hillenbrand97}, combined with
evolutionary tracks of \citet{Siess00}, gives
self-consistent ages for our sample
 with 481 stars of which 98.5\% are detected in the
COUP image.  These ages range from $\leq 0.1$~Myr to $\simeq 10$ Myr 
which covers a wide range of the PMS phases of low-mass stellar 
evolution.  The large sample permits us to establish the evolution 
of X-ray luminosities (and the related activity indicators $L_{\rm 
X}/L_{\rm bol}$ and $F_{\rm X}$) with higher precision and 
reliability than possible from past studies. In particular, we 
examine activity evolution in mass strata to avoid the convolution 
of temporal and mass dependencies of magnetic activity that have 
confused previous studies.  To establish the long-term evolution of 
magnetic activity, we compare our findings with mass-stratified XLFs 
obtained with the $ROSAT$ satellite for older stars, particularly 
the Pleiades, Hyades, and a volume-limited sample of solar 
neighborhood stars. 

Our fundamental result is that, treating the PMS phases as a whole, 
the X-ray luminosity decay law for stars in the $0.5 < M < 
1.2$~M$_\odot$ mass range is approximately powerlaw with $L_{\rm X} 
\propto \tau^{-0.75}$ over the wide range of ages $5 < \log \tau < 
9.5$ yr.  Studies of old disk stars suggest that this may 
steepen to $L_{\rm X} \propto \tau^{-1.5}$ over $9.0 < \tau < 
10.0$ yr \citep{Guedel97, Feigelson04}.  We also note that 
\citet{Pace04} find evidence for a very steep decay of the 
chromospheric activity between $8.7 < \tau < 9.3$ yr. These results 
are similar to, but show more rapid decay than, the classical 
\citet{Skumanich72} $\tau^{-1/2}$ relation which had been measured 
for main sequence stars only over the limited age range $7.5 < \log 
\tau < 9.5$ yr. We also establish for the first time a mild decay 
in magnetic activity roughly as $L_{\rm X} \propto \tau^{-1/3}$, 
for ages $0.1 < \tau < 10$ Myr within the PMS phase.

The magnetic decay is somewhat more complicated when other X-ray 
indicators and behavior within the PMS regime are considered. We 
find that the fractional X-ray luminosity $L_{\rm X}/L_{\rm bol}$ 
and X-ray surface flux $F_{\rm X}$ exhibit an accelerating decay 
with age, falling slowly during the $1-100$ Myr interval and 
faster on Gyr timescales on the main sequence.

While these results would appear to confirm and elaborate the 
long-standing rotation-age-activity relationship of solar-type 
stars, other recent $Chandra$ studies (cf. \S~\ref{intro.sec} and 
Preibisch et al.\ 2005) show that the rotation-activity relation is 
completely absent in PMS stars.  We thus find the somewhat 
surprising result that the activity-age decay is strong across the 
entire history of solar-type stars but is not entirely attributable 
to rotational deceleration.  A more complex astrophysical situation, 
perhaps involving both tachoclinal and convective dynamos as 
described by \citet{Barnes03a, Barnes03b}, is needed.  Within the 
PMS phase when the star is still contracting, changes in the 
convective volume and surface area may also be involved in the 
activity decay.

We furthermore show that the magnetic activity history for M 
stars with masses 0.1 to 0.4~M$_\odot$  appears to be different. 
Only a mild decrease in X-ray luminosity, and even a mild increase 
in $L_{\rm X}/L_{\rm bol}$ and $F_{\rm X}$, is seen over the $1-100$ 
Myr range, though the X-ray emission does decay over long Gyr on the 
main sequence. Although this result may be related to the 
well-established fact that the low-mass M stars have much longer 
rotational slow-down times than solar-type stars 
\citep[e.g.,][]{Stauffer86}, the non-existence of an 
activity-rotation relation for the $\sim 0.1-10$~Myr old ONC stars 
suggests that other astrophysical processes underlying the magnetic 
activity are responsible for this difference. Both PMS and main 
sequence dM stars are mostly or fully convective, and the unchanging 
fractional X-ray luminosity may represent the saturation of the 
magnetic dynamo processes in these convective stars. The difference 
in behavior compared to higher mass stars supports the idea that 
the dynamos in PMS and dM stars may be qualitatively different than 
in solar-type stars, arising from a dynamo powered by turbulence 
distributed throughout the convective interior rather than from 
rotational shear at the tachocline. These issues are further 
discussed in \citet{Mullan01}, \citet{Feigelson03}, 
\citet{Barnes03b}, and \citet{Preibisch05}. 

While a comprehensive interpretation of our results in terms of 
dynamo theory is beyond the scope of this study, we believe they 
provide basic support for the idea that distributed convective 
dynamos are the dominant source of magnetic fields in PMS and dM 
stars. 

\acknowledgements COUP is supported by $Chandra$ Guest Observer 
grant SAO GO3-4009A (E. Feigelson, PI). We would like to thank  
J.H.M.M.~Schmitt and C.~Liefke for information on the NEXXUS 
database, B.~Stelzer and E.~Flaccomio for X-ray data on young 
clusters, G.~Micela and J.H.~Kastner for thoughtful comments on 
the manuscript, and
the referee, Manuel Guedel, for his detailed, insightful, and
helpful comments on this paper. 

Facility: CXO(ACIS)

\bibliography{aj-jour}

\clearpage
\newpage

\clearpage
\newpage

\clearpage
\newpage

\clearpage
\newpage

\clearpage
\newpage

\clearpage
\newpage

\begin{deluxetable}{rrrr}
\centering \tabletypesize{\normalsize} \tablewidth{0pt}
\tablecolumns{4}

\tablecaption{X-ray activity versus isochronal age correlations 
for the ONC stars
\label{fit.tab}}

\tablehead{
\colhead{Mass range} & \colhead{$P(0)$} & \colhead{$a$} & \colhead{$b$} \\
$[M_\odot]$ &  &  & 
}
\startdata
\multicolumn{4}{c}{(a) $X = L_{\rm X}$}\\
$0.1-0.2$ &  0.021 &$30.31\pm 0.57$&$-0.22\pm  0.09 $\\
$0.2-0.4$ &  0.018 &$30.52\pm 0.48$&$-0.17\pm  0.08 $\\
$0.4-1.0$ &  0.001 &$32.17\pm 0.60$&$-0.32\pm  0.10 $\\
$1.0-2.0$ &  0.053 &$33.66\pm 1.39$&$-0.50\pm  0.21 $\\ 
\multicolumn{4}{c}{(b) $X = L_{\rm X}/L_{\rm bol}$} \\
$0.1-0.2$ &  0.030 &$-5.14\pm  0.53$&$  0.23\pm  0.09 $\\
$0.2-0.4$ &  0.001 &$-5.66\pm  0.48$&$  0.34\pm  0.08 $\\
$0.4-1.0$ &  0.010 &$-5.01\pm  0.57$&$  0.28\pm  0.09 $\\
$1.0-2.0$ &  0.789 &$-3.35\pm  1.62$&$ -0.02\pm  0.25 $\\ 
\multicolumn{4}{c}{(c) $X = F_{\rm X}$}\\
$0.1-0.2$ &  0.008 &$ 4.33\pm 0.53 $&$ 0.28\pm 0.09 $\\
$0.2-0.4$ &  0.001 &$ 4.02\pm 0.48 $&$ 0.37\pm 0.08 $\\
$0.4-1.0$ &  0.008 &$ 4.86\pm 0.57 $&$ 0.31\pm 0.09 $\\
$1.0-2.0$ &  0.499 &$ 6.08\pm 1.45 $&$ 0.13\pm 0.22 $\\
\enddata
\tablecomments{This table summarizes the
results of the correlation analysis for the relations
between X-ray activity and isochronal age of the
form $\log\left(X\right) = a + b \times \log\left(\tau\right)$.
We list the probabilities $P(0)$ that no correlation is present
found with the generalized nonparametric Kendall's tau
statistic and the coefficients of the linear regressions
calculated using the
EM Algorithm in ASURV.
}
\end{deluxetable}

\end{document}